**Malach Obisa Amonga**

**Department of computer Science**

**Chuka University**

# TITLE: A LITERATURE REVIEW OF RECENT ADVANCES IN SOFTWARE DESIGN AND ARCHITECTURE

## Abstract

Software architecture has evolved considerably in response to the increasing complexity of modern software systems, particularly those based on cloud computing, microservices, artificial intelligence (AI), and distributed computing environments. This literature review synthesizes recent studies published between 2024 and 2025 to examine emerging trends, challenges, and future directions in software design and architecture. The review adopts a thematic synthesis approach to analyse contemporary research across five major areas: architectural modelling and representation, software quality attributes and self-adaptive architectures, architectural evolution and complexity management, artificial intelligence-assisted architectural decision-making, and existing research gaps. The findings indicate that modern software architecture extends beyond traditional structural design to support continuous architectural governance, stakeholder communication, runtime observability, resilience, and intelligent decision support throughout the software lifecycle. Furthermore, the reviewed studies demonstrate that multiple architectural views, continuous monitoring, domain-driven decomposition, and AI-assisted design techniques contribute significantly to improving scalability, maintainability, adaptability, and long-term software sustainability. Despite these advances, several research gaps remain, including limited empirical validation of proposed approaches, insufficient integration of security and privacy into architectural decision-making, inadequate exploration of emerging paradigms such as edge and serverless computing, and the absence of standardized frameworks for trustworthy AI-assisted architecture. The review concludes that future software architecture will increasingly rely on the integration of intelligent automation with human expertise to address growing system complexity while maintaining alignment with organizational objectives and quality requirements. These findings provide a comprehensive overview of current research and identify promising directions for future investigation in software design and architecture.

## LITERATURE REVIEW

### Introduction

Software architecture has become a fundamental discipline in software engineering because it provides the structural framework that guides the design, implementation, deployment, maintenance, and evolution of software systems. As software applications continue to grow in size, complexity, and distribution, traditional architectural approaches have become increasingly inadequate for addressing the challenges associated with cloud computing, micro services, artificial intelligence, and continuously evolving business requirements. Consequently, recent research has shifted towards architectural approaches that promote scalability, adaptability, maintainability, resilience, and intelligent decision-making. The literature further demonstrates that software architecture is no longer regarded merely as a high-level design activity but as a continuous process that supports communication among stakeholders, runtime management, quality assurance, and long-term system evolution.

The selected studies published between 2024 and 2025 collectively examine several emerging issues in software architecture, including architectural modelling, architectural views, software quality attributes, self-adaptive systems, architectural evolution, complexity management, and artificial intelligence-assisted architectural decision-making. Although each study investigates a different aspect of software architecture, they collectively demonstrate the changing nature of architectural design in response to modern distributed software environments. Rather than considering each study independently, this review synthesizes their findings to demonstrate how contemporary software architecture research is converging towards more adaptive, intelligent, and maintainable architectural solutions.

### Architectural Modelling and Architectural Representation

One of the dominant themes emerging from recent software architecture research is the increasing need for effective architectural modelling techniques capable of representing highly distributed software systems. The rapid adoption of microservice architectures has fundamentally transformed software development by replacing large monolithic applications with collections of loosely coupled services. While this architectural style improves scalability, flexibility, and independent deployment, it simultaneously introduces considerable architectural complexity. Existing literature suggests that as the number of services increases, software architects experience growing difficulties in documenting architectural decisions, tracing dependencies among services, and maintaining consistency throughout the software lifecycle.

Brambilla et al. (2024) argue that conventional architectural modelling techniques were originally developed for relatively static software systems and therefore struggle to adequately represent the dynamic behaviour exhibited by modern microservice architectures. According to the authors, distributed systems continuously evolve through service additions, independent deployments, runtime scaling, and changing communication patterns, making traditional static architectural models increasingly insufficient. The study therefore proposes a modelling framework capable of integrating multiple architectural perspectives, including conceptual, logical, execution, and deployment views, thereby improving traceability between architectural decisions and implementation artefacts. Their findings suggest that architectural modelling should no longer be viewed simply as documentation but rather as an active mechanism for controlling complexity throughout software development.

The importance of multiple architectural representations is similarly emphasized by López et al. (2024), who argue that software architecture performs not only a technical function but also an important communication role among project stakeholders. The authors observe that software development involves diverse participants, including software architects, developers, project managers, customers, operations teams, and regulatory agencies, each requiring different architectural information to perform their responsibilities effectively. Consequently, they propose an architectural view model specifically designed for microservice-based systems that separates functional, development, runtime, and deployment concerns while maintaining consistency across the various architectural perspectives.

The arguments presented by Brambilla et al. (2024) and López et al. (2024) are highly complementary because both studies recognize that modern software architecture must accommodate increasing system complexity through structured architectural representations. While Brambilla et al. primarily focus on improving architectural modelling techniques for representing distributed software systems, López et al. extend this discussion by demonstrating how multiple architectural views facilitate stakeholder communication and support architectural decision-making. Collectively, these studies reinforce the principle that architecture serves not only as a technical blueprint but also as an organizational communication mechanism that enables different stakeholders to understand complex software systems from perspectives relevant to their specific responsibilities.

The reviewed studies further indicate that maintaining consistency across multiple architectural views has become increasingly important as organizations adopt DevOps practices, continuous integration, and cloud-native deployment strategies. Modern software architectures require documentation approaches capable of evolving alongside continuously changing implementations without introducing architectural drift. Consequently, architectural modelling is increasingly regarded as an integral component of software governance rather than a one-time design activity performed during project initiation.

**Software Quality Attributes, Runtime Reliability and Self-Adaptive Architectures**

Beyond architectural representation, contemporary research increasingly emphasizes software quality attributes as central considerations during architectural design. Traditional software architecture primarily focused on satisfying functional requirements; however, recent literature demonstrates that non-functional requirements, including reliability, scalability, maintainability, adaptability, availability, security, and performance, now play an equally important role in determining architectural success. This shift reflects the growing operational demands placed upon distributed software systems that must function continuously under changing environmental conditions.

The concept of adaptability receives significant attention in the work of Tavares and Rosa (2024), who examine the emergence of Human-in-the-Loop (HITL) architectures within self-adaptive software systems. The authors argue that modern software increasingly operates within environments characterized by uncertainty, fluctuating workloads, evolving cybersecurity threats, and continuously changing user requirements. Under such conditions, architectures designed around fixed assumptions often struggle to maintain acceptable levels of performance and reliability. Their review therefore advocates integrating automated adaptation mechanisms with human expertise, enabling software systems to respond intelligently while preserving accountability and contextual decision-making.

According to Tavares and Rosa (2024), effective self-adaptive architectures rely upon continuous monitoring, feedback loops, context-awareness mechanisms, and intelligent decision-support frameworks that enable systems to detect environmental changes and modify their behaviour accordingly. Nevertheless, the authors caution against complete automation, arguing that certain architectural decisions require human judgement, ethical consideration, and contextual understanding that cannot easily be replicated by autonomous systems.

The significance of runtime reliability is similarly emphasized by the MicroIRC study (2024), which investigates fault diagnosis within distributed microservice environments. Existing literature consistently identifies reliability as one of the most important software quality attributes because failures within interconnected services can rapidly propagate throughout an entire application. Unlike traditional monolithic systems where faults may remain localized, distributed microservices involve numerous interacting components whose dependencies complicate fault identification and recovery.

The MicroIRC framework addresses this challenge by introducing instance-level root cause localization capable of identifying failures at the individual service instance rather than only at the service level. This approach significantly improves architectural observability by enabling more precise monitoring, diagnosis, and recovery of runtime failures. The study therefore demonstrates that modern software architecture increasingly incorporates operational concerns traditionally associated with systems administration, including distributed tracing, logging, runtime monitoring, and resilience engineering.

Although Tavares and Rosa (2024) and the MicroIRC study investigate different architectural problems, both emphasize the growing importance of runtime intelligence within software systems. Tavares and Rosa primarily concentrate on adaptive decision-making through human-centred automation, whereas the MicroIRC research focuses on runtime reliability through intelligent fault localization. Together, these studies suggest that future software architectures will require continuous observation, intelligent adaptation, and resilient operational management to maintain acceptable levels of software quality in increasingly dynamic computing environments.

**Architectural Evolution, Complexity Management and Software Sustainability**

As software systems continue to evolve in response to changing business requirements, technological innovation, and user expectations, architectural evolution has become an increasingly important area of software architecture research. Unlike traditional software systems that remained relatively stable after deployment, modern cloud-native applications undergo continuous modification through frequent releases, feature enhancements, security updates, and infrastructure changes. Consequently, researchers increasingly argue that software architecture should be viewed as a living asset requiring continuous monitoring and governance throughout the software lifecycle rather than as a static blueprint developed only during the initial stages of system design.

The importance of architectural sustainability is extensively discussed by Abdelfattah et al. (2024), who investigate how microservice architectures evolve over time and how uncontrolled architectural changes may gradually reduce software quality. The authors observe that although microservices promote flexibility and independent service deployment, their long-term evolution often introduces unintended architectural consequences, including increased service dependencies, tighter coupling, architectural drift, and technical debt. Existing literature consistently recognizes architectural drift as one of the leading causes of declining software maintainability because incremental modifications gradually move implementation away from the intended architectural design.

To address these concerns, Abdelfattah et al. (2024) propose the application of static analysis techniques to evaluate architectural quality throughout software evolution. Their approach enables architects to continuously monitor communication patterns, dependency relationships, and compliance with architectural constraints before architectural degradation becomes difficult to reverse. The study therefore extends traditional software maintenance by emphasizing proactive architectural governance rather than reactive correction of software defects. This perspective suggests that architectural quality should be continuously assessed throughout development rather than only during design reviews.

The challenge of managing architectural complexity is further examined by Ciccozzi et al. (2025), who argue that many organizations adopting microservice architectures underestimate the complexity associated with defining appropriate service boundaries. Although decomposing applications into smaller services has become a widely accepted architectural strategy, the authors note that poor decomposition frequently produces excessive inter-service communication, duplicated functionality, inconsistent business logic, and unnecessary operational complexity. Consequently, simply dividing software into smaller components does not automatically produce a well-designed architecture.

Building upon Domain-Driven Design (DDD) principles, Ciccozzi et al. (2025) propose a multitree-based architectural framework that organizes services according to business domains and bounded contexts. Rather than structuring software solely around technical components, their approach aligns architectural decomposition with organizational business processes, thereby improving modularity, maintainability, scalability, and reuse. The authors argue that architectural decisions should reflect business requirements because software architecture ultimately exists to support organizational objectives rather than merely satisfy technical constraints.

Although Abdelfattah et al. (2024) and Ciccozzi et al. (2025) investigate different architectural problems, both studies converge on the broader objective of improving long-term architectural sustainability. Abdelfattah et al. focus primarily on monitoring architectural evolution and preventing architectural degradation, whereas Ciccozzi et al. emphasize the importance of designing architectures that minimize unnecessary complexity from the outset. Together, these studies suggest that achieving sustainable software architecture requires balancing continuous architectural governance with carefully planned architectural decomposition.

The reviewed literature also demonstrates that architectural evolution and complexity management are closely interconnected. Poor architectural decisions made during initial system design often become increasingly difficult and expensive to correct as software systems evolve. Consequently, recent studies advocate integrating architectural quality assessment into continuous software development practices, enabling organizations to detect emerging architectural problems early and maintain long-term software sustainability. This shift reflects the broader movement towards continuous architecture, where architectural decisions evolve alongside software implementation while preserving consistency with organizational goals.

## Artificial Intelligence and the Future of Software Architecture

Another important trend emerging from recent software architecture research is the increasing application of Artificial Intelligence (AI) to architectural design and decision-making. Advances in machine learning, generative AI, and intelligent automation have significantly influenced software engineering practices, prompting researchers to investigate how AI can support architects in analysing complex systems, evaluating design alternatives, predicting quality attributes, and managing architectural knowledge. As software systems become increasingly distributed and interconnected, manual architectural decision-making alone may no longer be sufficient to manage growing system complexity.

Bucaioni et al. (2025) provide one of the most comprehensive recent reviews examining the relationship between Artificial Intelligence and software architecture. According to the authors, software architects routinely encounter complex trade-offs involving performance, scalability, maintainability, security, availability, reliability, cost, and energy efficiency. Evaluating these competing quality attributes often requires extensive architectural experience, making architectural decision-making both time-consuming and cognitively demanding. Consequently, the authors argue that AI technologies have considerable potential to augment human expertise by providing intelligent recommendations based on accumulated architectural knowledge and historical design patterns.

The review identifies several architectural activities that can benefit from AI-assisted techniques, including architectural pattern selection, software quality prediction, automated documentation generation, architectural compliance verification, design recommendation, and architectural knowledge management. These capabilities suggest that AI may significantly reduce the effort associated with analysing complex software systems while improving the consistency and quality of architectural decisions. Furthermore, AI-supported tools have the potential to facilitate early identification of architectural risks before implementation, thereby reducing software maintenance costs and improving overall software quality.

Despite these potential benefits, Bucaioni et al. (2025) caution that Artificial Intelligence should complement rather than replace human architects. The reviewed literature indicates that software architecture frequently involves organizational objectives, ethical considerations, regulatory requirements, financial constraints, and contextual knowledge that cannot be adequately captured by automated reasoning alone. Consequently, architectural decision-making remains fundamentally a collaborative process in which AI functions as an intelligent decision-support mechanism rather than an autonomous architect.

The arguments presented by Bucaioni et al. (2025) complement the findings of the other reviewed studies. For example, the architectural modelling techniques proposed by Brambilla et al. (2024) could potentially benefit from AI-assisted model generation, while the architectural view models discussed by López et al. (2024) could be enhanced through automated consistency verification. Similarly, AI technologies may strengthen self-adaptive systems by improving runtime decision-making, support architectural evolution through predictive maintenance analysis, and assist complexity management by identifying architectural anti-patterns during system development. These connections demonstrate that Artificial Intelligence is not an isolated research area but rather an enabling technology capable of enhancing numerous aspects of software architecture.

Collectively, the literature suggests that AI will play an increasingly significant role in future software architecture by improving architectural analysis, facilitating knowledge reuse, supporting continuous architecture, and assisting software architects in managing increasingly complex distributed systems. However, the reviewed studies consistently emphasize that successful AI adoption requires maintaining appropriate human oversight to ensure that architectural decisions remain aligned with organizational objectives and stakeholder expectations.

## Research Gaps in the Reviewed Literature

Although the reviewed studies make significant contributions to contemporary software architecture, several research gaps remain evident. First, most of the studies concentrate primarily on microservice architectures, leaving comparatively limited attention to other emerging architectural paradigms such as serverless computing, edge computing, cyber-physical systems, and Internet of Things (IoT) architectures. As software systems continue to diversify, future research should investigate whether the proposed architectural models remain applicable across different computing environments.

Secondly, while several studies discuss architectural quality attributes such as maintainability, scalability, reliability, and adaptability, relatively few provide comprehensive empirical evaluations comparing the effectiveness of competing architectural approaches under realistic industrial conditions. Most contributions introduce frameworks, conceptual models, or surveys without extensive longitudinal validation across multiple application domains. Additional empirical studies involving large-scale industrial software systems would therefore strengthen the practical applicability of existing architectural recommendations.

Another noticeable gap concerns the integration of security and privacy considerations within architectural decision-making. Although reliability and adaptability receive substantial attention throughout the reviewed literature, cybersecurity is often addressed indirectly rather than being incorporated as a primary architectural concern. Considering the increasing prevalence of cyber threats targeting distributed cloud-native applications, future architectural research should place greater emphasis on secure-by-design architectural principles.

Furthermore, despite the growing interest in Artificial Intelligence for software architecture, current research remains largely exploratory. Existing studies primarily discuss potential applications of AI rather than providing standardized methodologies for integrating AI into everyday architectural practice. Consequently, further investigation is required to establish practical frameworks, evaluation metrics, governance models, and ethical guidelines that ensure trustworthy and explainable AI-assisted software architecture.

Finally, the reviewed literature demonstrates limited discussion regarding sustainability and green software architecture. As organizations become increasingly concerned with energy consumption and environmental sustainability, future architectural research should explore approaches that simultaneously optimize software quality, computational efficiency, and environmental impact.

## Conclusion

The reviewed literature demonstrates that software architecture has undergone substantial transformation in response to the growing complexity of contemporary software systems. Collectively, the studies indicate that software architecture now extends beyond traditional system design to encompass architectural modelling, stakeholder communication, runtime reliability, self-adaptation, architectural governance, complexity management, and intelligent decision support. This evolution reflects the changing role of software architecture as an ongoing organizational capability that supports software development throughout the entire system lifecycle.

The literature consistently demonstrates that effective architectural modelling enhances software understanding and communication, while multiple architectural views improve collaboration among diverse stakeholders. Similarly, the reviewed studies emphasize that software quality increasingly depends upon runtime observability, adaptive behaviour, and continuous architectural monitoring rather than solely on initial design decisions. The research also illustrates that sustainable software architecture requires continuous evaluation of architectural evolution, careful management of complexity, and alignment between software structures and business domains.

Perhaps the most significant trend emerging from the reviewed literature is the increasing integration of Artificial Intelligence into software architecture. Rather than replacing software architects, AI is expected to enhance architectural analysis, improve decision support, facilitate architectural knowledge management, and strengthen continuous architectural governance. Nevertheless, human expertise remains essential for balancing technical decisions with organizational objectives, ethical considerations, and stakeholder expectations.

## References

Abdelfattah, A. S., Cerny, T., Bushong, V., Al Maruf, A., & Taibi, D. (2024). *Assessing evolution of microservices using static analysis*. Applied Sciences, 14(22), 10725. https://doi.org/10.3390/app142210725

Brambilla, D., Esparza Peidro, J., Muñoz-Escoí, F. D., & Bernabéu Auban, J. M. (2024). *Modeling microservice architectures*. Journal of Systems and Software, 213, 112041. https://doi.org/10.1016/j.jss.2024.112041

Bucaioni, A., Weyssow, M., He, J., Lyu, Y., & Lo, D. (2025). *Artificial intelligence for software architecture: Literature review and the road ahead*. arXiv. https://arxiv.org/abs/2504.04334

Ciccozzi, F., et al. (2025). *Layered microservices architecture: A multitree-based domain-driven approach*. Information and Software Technology, 181, 107720. https://doi.org/10.1016/j.infsof.2025.107720

López, J. A., Huynh Anh, V. N., et al. (2024). *An architectural view model for designing and implementing microservices-based systems: Use case in FinTech*. Procedia Computer Science, 237, 667–674. https://doi.org/10.1016/j.procs.2024.05.152

Tavares, G. J. S., & Rosa, N. S. (2024). *A survey on human in the loop for self-adaptive systems*. Journal of Universal Computer Science, 30(11). https://doi.org/10.3897/jucs.114513

Zhu, Y., et al. (2024). *MicroIRC: Instance-level root cause localization for microservice systems*. Journal of Systems and Software.